\documentclass[manuscript]{aastex}
\usepackage{amssymb}

\slugcomment{To be published in ApJ}
\shorttitle{Oblique Alfv\'{e}n instability}
\shortauthors{Malovichko et al.}

\begin{document}

\title{Oblique Alfv\'{e}n instability driven by compensated currents}

\author{P. Malovichko}
\affil{Main Astronomical Observatory, NASU, Kyiv, Ukraine}

\author{Y. Voitenko}
\affil{Solar-Terrestrial Centre of Excellence, Space Physics Division,
Belgian Institute for Space Aeronomy, Ringlaan-3-Avenue Circulaire,
B-1180 Brussels, Belgium}
\email{voitenko@oma.be}

\and

\author{J. De Keyser}
\affil{Solar-Terrestrial Centre of Excellence, Space Physics Division,
Belgian Institute for Space Aeronomy, Ringlaan-3-Avenue Circulaire,
B-1180 Brussels, Belgium}

\begin{abstract}
Compensated-current systems created by energetic ion beams are widespread
in space and astrophysical plasmas. The well-known examples are foreshock
regions in the solar wind and around supernova remnants. We found a new
oblique Alfv\'{e}nic instability driven by compensated currents flowing
along the background magnetic field. Because of the vastly different
electron and ion gyroradii, oblique Alfv\'{e}nic perturbations react
differently on the currents carried by the hot ion beams and the return
electron currents. Ultimately, this difference leads to a non-resonant
aperiodic instability at perpendicular wavelengths close to the beam ion
gyroradius. The instability growth rate increases with increasing beam
current and temperature. In the solar wind upstream of Earth's bow shock
the instability growth time can drop below 10 proton cyclotron periods.
Our results suggest that this instability can contribute to the turbulence
and ion acceleration in space and astrophysical foreshocks.
\end{abstract}


\section{Introduction}

Current-driven instabilities are important for anomalous resistivity and
related energy release in weakly collisional space plasmas, like the solar
corona, solar wind and planetary magnetospheres. Kinetic instabilities of
ion-acoustic, ion-cyclotron, and lower-hybrid drift waves were studied
extensively in this context (Duijveman et al., 1981; B\"{u}chner \&
Elkina 2006, and references therein). More recently, the electron-current
driven kinetic Alfv\'{e}n instability has been discussed as a possible
source for the anomalous resistivity (Voitenko, 1995), and the anomalous
resistivity scaling in solar flares has been shown to be compatible with the
kinetic Alfv\'{e}n scenario (Singh \& Subramanian, 2007). These
instabilities were classified as resonant instabilities driven by the
inverse electron Landau damping (i.e., by the Cherenkov-resonant electrons).
The off-resonant electrons with velocities far from the wave phase
velocities interact with waves adiabatically and do not contribute to the
wave growth.

On the contrary, the non-resonant current instability of the Alfv\'{e}n mode
(Malovichko \& Iukhimuk, 1992a,1992b; Malovichko, 2007) is driven by the total
electric current rather than the resonant electrons only. This "pure"
current instability (PCI) was originally applied to the terrestrial auroral
zones (Malovichko \& Iukhimuk, 1992a), and then studied in application to
the coronal loops in the solar atmosphere (Malovichko \& Iukhimuk, 1992b;
Malovichko, 2007; Chen \& Wu, 2012). PCI instability appeared to be
universal in sense that the threshold current is virtually zero in uniform
unbounded plasmas.

Besides the applied external electric fields (electrostatic of inductive),
the electron and ion currents can also be induced in the background plasma
by other sources, like injected particle beams. We are interested here in
the cases where the currents injected by hot ion beams are compensated by
the return currents of the background electrons (or by co-propagating
electron beams). This situation occurs in many space and astrophysical
plasmas. For example, high-energy ion beams accelerated by shocks set up
compensated-current systems upstream of the terrestrial bow shock (Paschmann
et al., 1981), and references therein) and around supernova remnants (Bell,
2005, and references therein).

In addition to the mentioned above current driven instabilities, several
ion-beam instabilities can develop in the compensated-current systems
created by the ion beams. In the early works by Sentmann et al. (1981),
Winske \& Leroy (1984), Gary (1985), it was believed that the
parallel-propagating Alfv\'{e}n and fast modes are most unstable. Later on,
the resonant oblique (kinetic) Alfv\'{e}n instabilities driven by ion beams
have been shown to be more important in certain parameter ranges. So,
analytical treatments (Voitenko, 1998; Voitenko \& Goossens, 2003;
Verscharen \& Chandran, 2013) and numerical simulations (Daughton et al.
1999; Gary et al. 2000) have demonstrated that the ion-cyclotron (Alfv\'{e}n
I in the terminology by Daughton et al.) and Cherenkov (Alfv\'{e}n II)
instabilities of oblique Alfv\'{e}n waves are often faster than the parallel
ones. These instabilities were in particular studied in application to the
alpha-particle flows in the solar wind (Gary et al., 2000; Verscharen \&
Chandran 2013).

Because of the incomplete knowledge of plasma instabilities that can arise,
behavior of such complex systems is still not well understood. For example,
evolution of the ions reflected from the terrestrial bow shock, and
responsible waves and instabilities, remain uncertain. The same concerns
cosmic ray acceleration by the shocks around super-nova remnants. It is
important to know what instabilities can arise there, and which one can
dominate for particular beam and plasma parameters. Recently, a new MHD-type
instability driven by the return currents induced by cosmic rays in the
foreshock plasma around supernova remnants was found by Bell (2004,2005). A
similar return-current instability was found earlier by Winske \& Leroy
(1983), but they did not elucidate the main physical factor leading to the
instability and hence did not categorize it as a current-driven.

In the present paper we investigate a new non-resonant instability that
arises in the compensated-current systems created by fast and hot ion beams.
In such systems the beam current is compensated by the return background
current and one should not expect PCI studied by Malovichko \& Iukhimuk and
by Chen \& Wu. However, oblique Alfv\'{e}n perturbations with short
perpendicular wavelengths respond differently to the currents carried by the
electrons and the currents carried by the ions. The difference arises
because of the different ion and electron gyroradii, such that the ion and
electron current-related terms do not cancel each other. The resulting
compensated-current oblique instability (CCOI) develops at sufficiently high
beam currents and temperatures. The instability is essentially oblique and
its growth rate attains a maximum when its cross-field wavelength is close
to the beam ion gyroradius.

\section{Problem setup}

A particular compensated-current system is considered consisting of the
low-density hot ion beam propagating along $\mathbf{B}_{0}$, the motionless
background ions, and the electron components providing the neutralizing
return current. The neutralizing electron current can be set up by the
background electron component and/or by the co-propagating electron beam. In
the context of our study it is important to note that the final result does
not depend on the way how the return electron current is set up; it is
enough that the gyroradius of the current-carrying electrons is much smaller
than the ion beam gyroradius.

For each unperturb plasma component we use a $u_{\alpha }$-shifted
Maxwellian velocity distribution
\begin{equation}
f_{0\alpha }=\frac{n_{\alpha }}{(2\pi T_{\alpha }/m_{\alpha })^{3/2}}\exp
\left( -\frac{m_{\alpha }v_{\bot }^{2}}{2T_{\alpha }}-\frac{m_{\alpha
}(v_{z}-u_{\alpha })^{2}}{2T_{\alpha }}\right) ,  \label{f0}
\end{equation}%
where $n_{\alpha }$, $T_{\alpha }$, and\ $u_{\alpha }$\ are the mean number
density, temperature, and parallel velocity, respectively, and $m_{\alpha }$
is the particle mass. The species $\alpha $ can be background ions ($i$),
background electrons ($e$), beam ions ($b$), and beam electrons ($be$). The
subscripts $z$ and $\perp $ indicate directions parallel and perpendicular
to the mean magnetic field $\mathbf{B}_{0}$.

The plasma is assumed to be charge-neutral $\sum_{\alpha }q_{\alpha
}n_{\alpha }=0$ and current-neutral, $\sum_{\alpha }q_{\alpha }n_{\alpha
}u_{\alpha }=0$. In the reference system of background protons the zero net
current condition reads as
\begin{equation}
\sum_{e}n_{e}u_{ze}=n_{b}u_{b},  \label{j=0}
\end{equation}%
where summation is over all electron components.

To study electromagnetic perturbations in such system we use a kinetic
plasma model, where the velocity distribution function of each specie $%
\alpha $ obeys the collisionless Vlasov equation
\begin{equation}
\frac{\partial f_{\alpha }}{\partial t}+\mathbf{v}\frac{\partial f_{\alpha }%
}{\partial \mathbf{r}}+\frac{q_{\alpha }}{m_{\alpha }}\left( \mathbf{E}+%
\frac{1}{c}\left[ \mathbf{v\times B}\right] \right) \frac{\partial f_{\alpha
}}{\partial \mathbf{v}}=0.  \label{f}
\end{equation}%
The self-consistent electric $\mathbf{E}$ and magnetic $\mathbf{B}$ fields
obey Maxwell equations with the charge density $\sum_{\alpha }q_{\alpha
}\int d^{3}vf_{\alpha }$ and the current density $\sum_{\alpha }q_{\alpha
}\int d^{3}v\mathbf{v}f_{\alpha }$, $q_{\alpha }$ and $m_{\alpha }$\ are the
particles charge and mass, $t$ - time, $\mathbf{r}$ - spatial coordinates,
and $\mathbf{v}$ - velocity-space coordinates.

\section{Low-frequency Alfv\'{e}nic solution}

Linearizing (\ref{f}) and Maxwell equations around unperturbed state ($%
f_{\alpha }=f_{0\alpha }+\delta f$, $\mathbf{E=E}_{0}\mathbf{+}\delta
\mathbf{E}$, $\mathbf{B=B}_{0}\mathbf{+}\delta \mathbf{B}$), one can reduce
the resulting linear Vlasov-Maxwell set of equations to three equations for
three components of the perturbed electric field $\delta E_{x}$, $\delta
E_{y}$, and $\delta E_{z}$. The nontrivial solutions to the Maxwell-Vlasov
set of equations, $\delta \mathbf{E}\neq 0$, exist if the wave frequency $%
\omega $ and the wave vector $\mathbf{k}=(k_{x},0,k_{z})$ satisfy the
following dispersion equation (e.g., Alexandrov, Bogdankevich, \& Rukhadze,
1984):
\begin{equation}
\left\vert k^{2}\delta _{ij}-k_{i}k_{j}-\frac{\omega ^{2}}{c^{2}}\varepsilon
_{ij}\right\vert =0,  \label{det}
\end{equation}%
where $\varepsilon _{ij}$ is the dielectric tensor, and $\delta _{ij}$ is
the Kronecker's delta-symbol. Using expressions for the elements $%
\varepsilon _{ij}$ given by (Alexandrov et al., 1984), we reduced them in
the low-frequency domain ($\left( \omega ^{\prime }/\omega _{Bi}\right)
^{2}\ll 1$ and $\left( k_{z}V_{T\alpha }/\omega _{B\alpha }\right) ^{2}\ll 1$%
) as follows:
\begin{eqnarray}
\varepsilon _{xx} &\simeq &1+\sum_{\alpha }\left( \frac{\omega _{P\alpha }}{%
\omega _{B\alpha }}\right) ^{2}\left( \frac{\omega _{\alpha }^{\prime }}{%
\omega }\right) ^{2}\frac{\left( 1-A_{0}\left( \mu _{\alpha }^{2}\right)
\right) }{\mu _{\alpha }^{2}};  \nonumber \\
\varepsilon _{xy} &=&-\varepsilon _{yx}\simeq -i\sum_{\alpha }\left( \frac{%
\omega _{P\alpha }^{2}}{\omega _{B\alpha }}\right) \left( \frac{\omega
_{\alpha }^{^{\prime }}}{\omega ^{2}}\right) A_{0}^{\prime }\left( \mu
_{\alpha }^{2}\right) ;  \nonumber \\
\varepsilon _{xz} &=&\varepsilon _{zx}\simeq \sum_{\alpha }\left( \frac{%
\omega _{P\alpha }}{\omega _{B\alpha }}\right) ^{2}\left( \frac{\omega
_{\alpha }^{^{\prime }}k_{x}u_{\alpha }}{\omega ^{2}}\right) \frac{\left(
1-A_{0}\left( \mu _{\alpha }^{2}\right) \right) }{\mu _{\alpha }^{2}};
\nonumber \\
\varepsilon _{yy} &\simeq &\varepsilon _{xx}+2\sum_{\alpha }\left( \frac{%
\omega _{P\alpha }}{\omega }\right) ^{2}\mu _{\alpha }^{2}A_{0}^{\prime
}(\mu _{\alpha }^{2})J_{+}\left( \xi _{\alpha }\right) ;  \nonumber \\
\varepsilon _{yz} &=&-\varepsilon _{zy}\simeq -i\frac{k_{x}}{k_{z}}%
\sum_{\alpha }\left( \frac{\omega _{P\alpha }^{2}}{\omega _{B\alpha }\omega }%
\right) A_{0}^{\prime }(\mu _{\alpha }^{2})\left[ 1-J_{+}\left( \xi _{\alpha
}\right) \right] ;  \nonumber \\
\varepsilon _{zz} &\simeq &1+\sum_{\alpha }\left( \frac{\omega _{P\alpha }}{%
k_{z}V_{T\alpha }}\right) ^{2}A_{0}\left( \mu _{\alpha }^{2}\right) \left[
1-J_{+}\left( \xi _{\alpha }\right) \right] +\sum_{\alpha }\left( \frac{%
\omega _{P\alpha }}{\omega _{B\alpha }}\right) ^{2}\left( \frac{%
k_{x}u_{\alpha }}{\omega }\right) ^{2}\frac{\left( 1-A_{0}\left( \mu
_{\alpha }^{2}\right) \right) }{\mu _{\alpha }^{2}},  \label{e}
\end{eqnarray}%
where $\omega _{\alpha }^{\prime }=\omega -k_{z}u_{\alpha }$,\ $A_{0}\left(
\mu _{\alpha }^{2}\right) =I_{0}\left( \mu _{\alpha }^{2}\right) \exp \left(
-\mu _{\alpha }^{2}\right) ,$ $I_{0}\left( \mu _{\alpha }^{2}\right) $ is
the zero-order modified Bessel function,\ $\mu _{\alpha }=k_{x}V_{T\alpha
}/\omega _{B\alpha }$ is the normalized perpendicular wavenumber, $%
A_{0}^{\prime }(x)=dA_{0}(x)/dx$, $\omega _{P\alpha }$\ ($\omega _{B\alpha }$%
) is the plasma (cyclotron) frequency, $V_{T\alpha }=\sqrt{T_{\alpha
}/m_{\alpha }}$\ is the thermal velocity. and $\xi _{\alpha }=\omega
_{\alpha }^{^{\prime }}/(k_{z}V_{T\alpha })$.

We found that the function $J_{+}\left( \xi _{\alpha }\right) $ (Alexandrov
et al., 1984),\
\begin{equation}
J_{+}\left( \xi _{\alpha }\right) =\xi _{\alpha }\exp \left( -\frac{\xi
_{\alpha }^{2}}{2}\right) \int_{i\infty }^{\xi _{\alpha }}dt\exp \left(
\frac{t^{2}}{2}\right) ,  \label{j}
\end{equation}%
is particularly useful in the context of present study. This function is
related to the well-known plasma $W$-function, $J_{+}\left( x\right) =-i%
\sqrt{\pi /2}xW\left( x/\sqrt{2}\right) $, and can be expanded in the small
and large argument series:
\begin{equation}
J_{+}\left( x\right) =x^{2}+O\left( x^{4}\right) -i\sqrt{\frac{\pi }{2}}%
x\exp \left( -\frac{x^{2}}{2}\right) ,\qquad \left\vert x\right\vert \ll 1;
\label{j<}
\end{equation}%
and
\begin{equation}
J_{+}\left( x\right) =1+\frac{1}{x^{2}}+O\left( \frac{1}{x^{4}}\right)
-i\eta \sqrt{\frac{\pi }{2}}x\exp \left( -\frac{x^{2}}{2}\right) ,\qquad
\left\vert x\right\vert \gg 1,  \label{j>}
\end{equation}%
where $\eta =0$ for $\mathrm{Im}x>0$, $\eta =1$ for $\mathrm{Im}x=0$, and $%
\eta =2$ for $\mathrm{Im}x<0$.

In low-$\beta $ plasmas, where the gas/magnetic pressure ratio of the
background plasma $\beta <1$, and for perpendicular wavelength smaller than
the background ion gyroradius, $\mu _{i}^{2}<<1$, the dispersion equation (%
\ref{det}) reduces to
\begin{equation}
\left( \omega ^{2}-k_{z}^{2}V_{A}^{2}\right) \left( \omega
^{2}-k^{2}V_{A}^{2}\right) \simeq \omega _{Bi}^{2}k_{z}^{2}V_{A}^{2}\left(
1+A_{0}^{\prime }(\mu _{b}^{2})\right) ^{2}\bar{j}_{b}^{2},  \label{DE}
\end{equation}%
where $V_{A}=B_{0}/\sqrt{4\pi n_{i}m_{i}}$ is the background Alfv\'{e}n
velocity, $n_{i}$\ is the background ion number density, $\bar{j}%
_{b}=n_{b}V_{b}/n_{i}V_{A}$ is the ion beam current $j_{b}=en_{b}V_{b}$
normalized by the Alfv\'{e}n current $j_{A}=en_{i}V_{A}$, and $V_{b}$ is the
mean beam velocity ($e$ is the proton charge and we assumed that the ions
are protons). When deriving (\ref{DE}) we used expansion (\ref{j<}) for the
background electron $J_{+}(\xi _{e})$, expansion (\ref{j>}) for the
background ion $J_{+}(\xi _{i})$, zero net current condition (\ref{j=0}),
and dropped small beam terms containing functions $J_{+}(\xi _{b})$ and $%
J_{+}(\xi _{be})$. Expansion (\ref{j>}) is already good for $J_{+}(\xi _{i})$
at $\xi _{i}\gtrsim 2$, where the relative error is less than 0.1. Note that
$\mu _{i}^{2}<<1$ does not imply $\mu _{b}^{2}=(T_{b}/T_{i})\mu _{i}^{2}<<1$
if the beam is hotter than the background.

In the absence of compensated currents, $\bar{j}_{b}=$ $j_{e}=$ $0$, the
equation (\ref{DE}) splits into two independent equations for MHD Alfv\'{e}n
and fast mode waves (the slow mode was dropped from (\ref{DE}) because of
the low plasma $\beta $). In the presence of $\bar{j}_{b}$ the Alfv\'{e}n
and fast mode waves are coupled, and the solution corresponding to Alfv\'{e}%
n mode reads as
\begin{equation}
\omega ^{2}=k_{z}^{2}V_{A}^{2}+\frac{1}{2}\left( k_{x}^{2}V_{A}^{2}-\sqrt{%
k_{x}^{4}V_{A}^{4}+4\omega _{Bi}^{2}k_{z}^{2}V_{A}^{2}\left( 1+A_{0}^{\prime
}(\mu _{b}^{2})\right) ^{2}\bar{j}_{b}^{2}}\right) .  \label{AW}
\end{equation}

The dispersion relation (\ref{AW}) for Alfv\'{e}nic perturbations provides a
basis for our analysis. The current term containing $\bar{j}_{b}$ shifts
down the Alfv\'{e}n wave frequency squared and can make it negative, which
means an aperiodic instability. The fast mode solution is up-shifted and
remains stable; we will not consider it here.

\section{Instability analysis}

\subsection{Threshold}

It is easy to see from (\ref{AW}) that the frequency becomes purely
imaginary, $\omega ^{2}<0$, when the beam current is sufficiently high,
\begin{equation}
\bar{j}_{b}>\frac{\mu _{b}}{\left( 1+A_{0}^{\prime }(\mu _{b}^{2})\right) }%
\left( \frac{k}{k_{x}}\right) \left( \frac{V_{A}}{V_{Tb}}\right) ,
\label{j>thr}
\end{equation}%
Function $\mu _{b}/\left( 1+A_{0}^{\prime }(\mu _{b}^{2})\right) \rightarrow
\infty $ in the limits $\mu _{b}\rightarrow 0$\ and $\mu _{b}\rightarrow
\infty $ and attains a minimum $\simeq 1.33$ at $\mu _{b}\simeq 0.89$. This
minimum defines the instability threshold current
\begin{equation}
\bar{j}_{\mathrm{thr}}\simeq 1.33\frac{V_{A}}{V_{Tb}}.  \label{jthr}
\end{equation}

Formally, the instability threshold is achieved at $k_{z}=0$. But in reality
$k_{z}$ has a lower bound defined by the parallel system scale, $%
k_{z}\gtrsim 2\pi /L_{z}$. This limitation does not alter the above
threshold estimation for realistic system length scales larger than the beam
ion gyroradius, $L_{z}\gg \rho _{b}$, $\rho _{b}=V_{Tb}/\omega _{Bi}$.

As an example, we compare the threshold of CCOI (\ref{jthr}) in the form
\begin{equation}
\left( \frac{n_{b}}{n_{i}}\right) _{\min }\left( \frac{V_{b}}{V_{A}}\right)
>1.33\left( \frac{V_{A}}{V_{Tb}}\right) ,  \label{nthr}
\end{equation}%
with the threshold of the fire-hose instability (FHI) in streaming plasmas
(e.g., Voitenko, Likhachev, \& Iukhimuk, 1980)
\begin{equation}
\left( \frac{n_{b}}{n_{i}}\right) _{\min }\left( \frac{V_{b}}{V_{A}}\right)
^{2}>1+\Delta ,  \label{nfh}
\end{equation}%
where anisotropy effects represented by $\Delta $ include plasma $\beta $
(temperature) anisotropies and heat flux ($q$) anisotropies:
\[
\Delta =\sum_{s=i,b}\left( \frac{\beta _{s\perp }-\beta _{s\parallel }}{2}%
\right) +\frac{8\pi k}{\omega _{Bi}B_{0}^{2}}\sum_{s=i,b}\left( q_{s\perp
}-q_{s\parallel }\right) .
\]%
Both $\Delta >0$ and $\Delta <0$ cases occur in the solar wind with $\Delta
>0$ dominated by anisotropic ion cores and $\Delta <0$ dominated by parallel
\textbf{electron strahls and halos, and by parallel ion tails and beams}
(Marsch 2006, and references therein). In general, the beam-driven fire-hose
threshold is reduced by $\Delta <0$ and increased by $\Delta >0$. We will
consider hereafter only the isotropic case $\Delta =0$, implying a pure
beam-driven fire-hose instability.

\begin{figure}[tbp]
\plotone{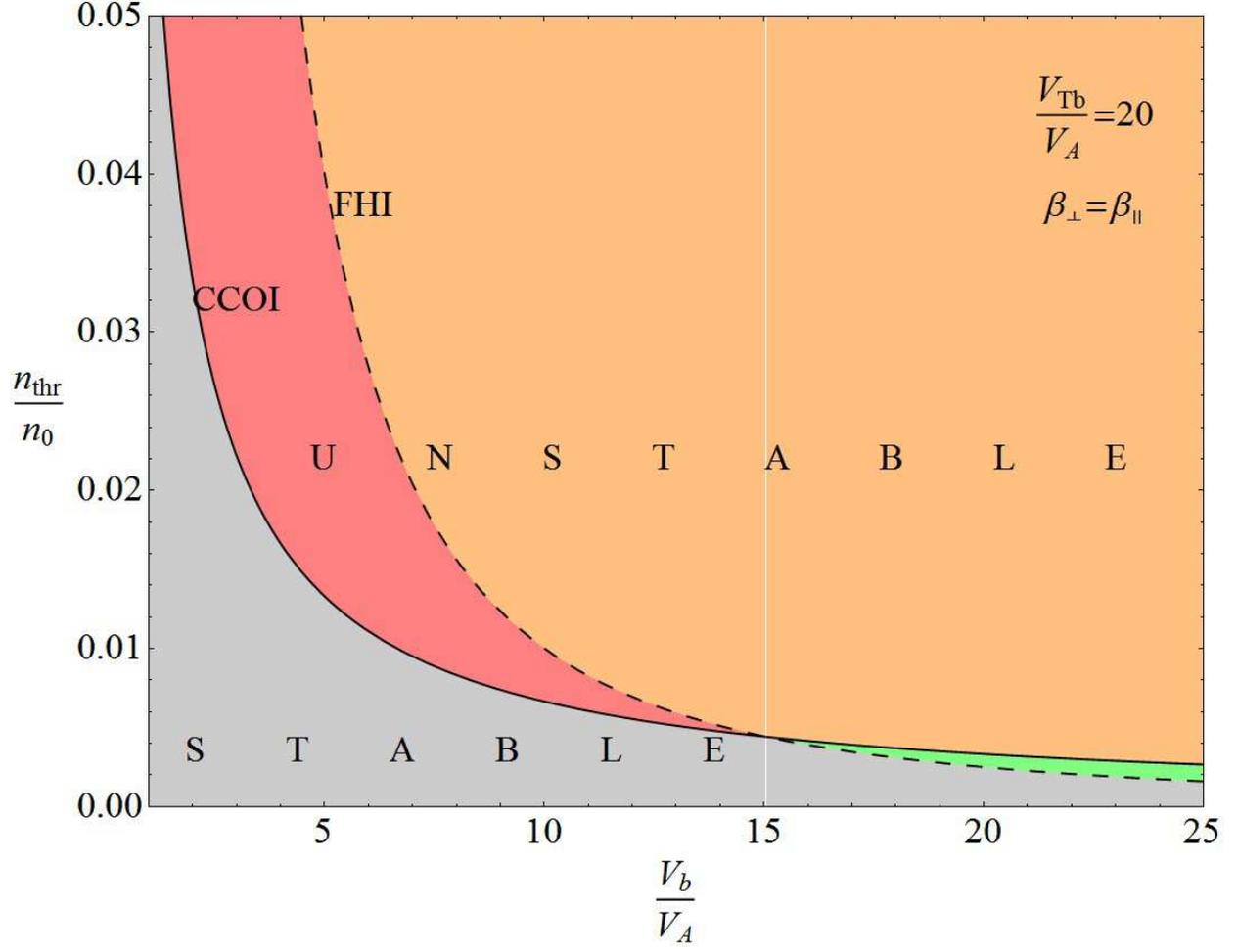}
\caption{(colour online) Comparison of the thresholds for the fire-hose (FH)
instability and CCOI. The beam thermal velocity is $V_{Tb}/V_{A}=20$ and the
plasma temperature is isotropic. The CCOI threshold is lower in the wide
range of beam velocities $V_{b}/V_{A}<15$. }
\end{figure}

The comparison of CCOI and FHI thresholds is shown in Fig. 1 for the case of
hot beam, $V_{Tb}/V_{A}=20$, in the isotropic background, $\Delta =0$. The
CCOI threshold for such hot beam is significantly lower than the FHI
threshold in a wide range of beam velocities. These parameter ranges are
relevant for compensated-current systems created in the solar wind by hot
ion beams propagating upstream of the terrestrial bow shock (see in more
detail below).

The parallel-propagating left- and right-hand resonant instabilities studied
by Gary (1985) have the velocity thresholds $V_{b\mathrm{thr}}^{-}/V_{A}\sim
0.82$ and $V_{b\mathrm{thr}}^{+}/V_{A}\sim 1$, respectively. The CCOI
velocity threshold found from (\ref{nthr}),
\[
\frac{V_{b\mathrm{thr}}^{CCOI}}{V_{A}}=1.33\left( \frac{n_{i}V_{A}}{%
n_{b}V_{Tb}}\right) ,
\]%
is lower than both thresholds of resonant instabilities $V_{b\mathrm{thr}%
}^{\pm }/V_{A}$ provided
\[
\frac{V_{Tb}}{V_{A}}>1.6\frac{n_{i}}{n_{b}}.
\]%
This condition is not easily satisfied in the terrestrial foreshock. Say,
for relatively high-density beam, $n_{b}/n_{i}\simeq 0.1$, it is satisfied
if the beam is also quite hot $V_{Tb}/V_{A}\geq 16$. However, as is shown
below, CCOI can be a strongest instability in the parameter range where
several instabilities are over-threshold.

\subsection{Wavenumber dependence of the instability increment}

To analyze the CCOI growth rate $\gamma =$Im$\left( \omega \right) $ as
function of wave vector components we rewrite (\ref{AW}) in the
dimensionless form:
\begin{equation}
\frac{\gamma }{\omega _{Bi}}=\sqrt{\sqrt{0.25\left( \frac{V_{A}}{V_{Tb}}%
\right) ^{4}\mu _{b}^{4}+\bar{j}_{b}^{2}\left( 1+A_{0}^{\prime }\left( \mu
_{b}^{2}\right) \right) ^{2}\left( \frac{k_{z}V_{A}}{\omega _{Bi}}\right)
^{2}}-\left( \frac{k_{z}V_{A}}{\omega _{Bi}}\right) ^{2}-0.5\mu
_{b}^{2}\left( \frac{V_{A}}{V_{Tb}}\right) ^{2}}.  \label{AW1}
\end{equation}%
We choose the normalization for the perpendicular wavenumber $\mu
_{b}=k_{\perp }\rho _{b}$, which simplifies the analysis of the current term
containing $A_{0}^{\prime }\left( \mu _{b}^{2}\right) $.

\begin{figure}[tbp]
\label{f2} \plotone{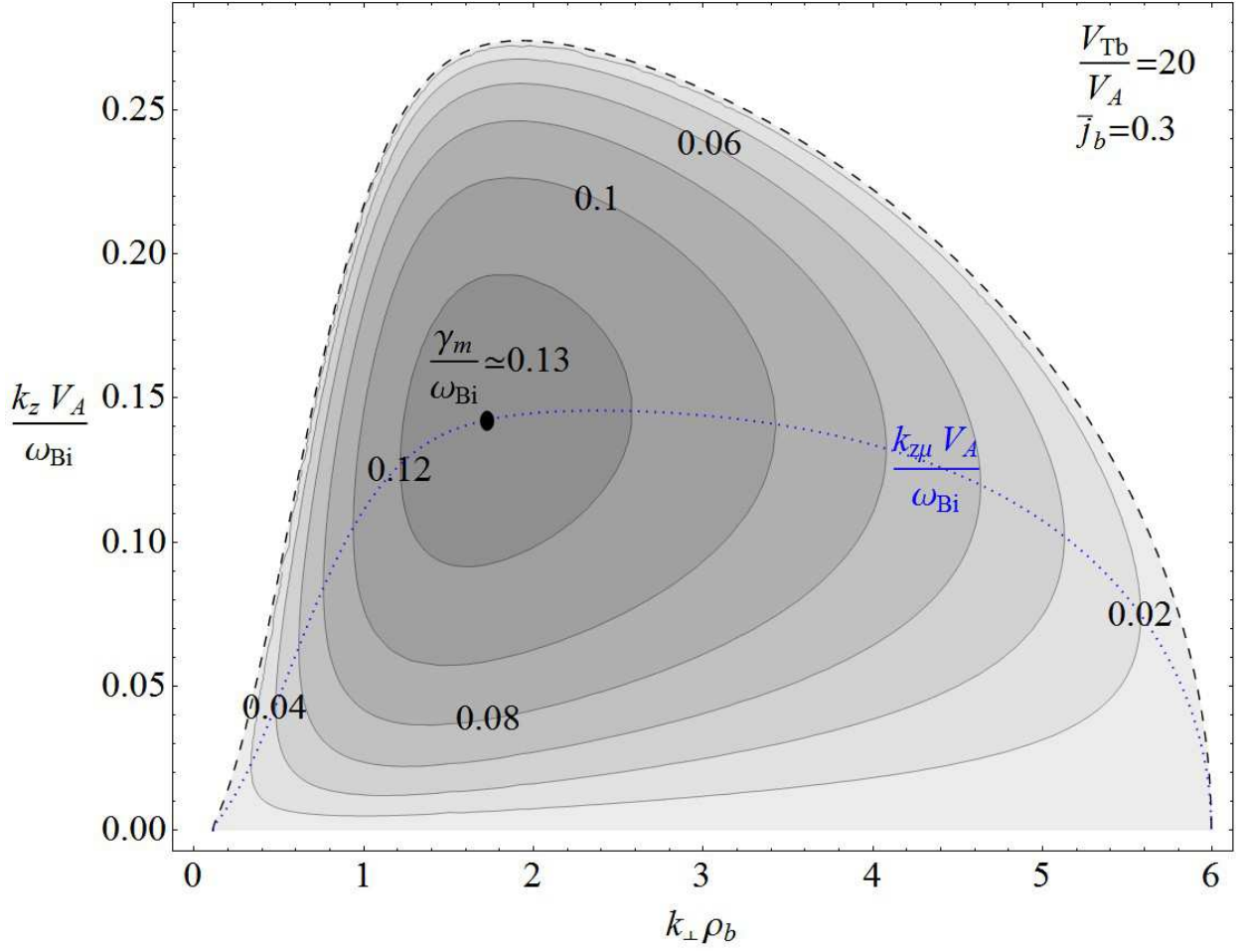}
\caption{(color online) Contours of the CCOI growth rate as function of
normalized parallel and perpendicular wavenumbers. The normalized beam
current $\bar{j}_{b}=0.3$ and thermal velocity $V_{Tb}/V_{A}=20$. The
absolute maximum of CCOI, $\protect\gamma _{\mathrm{m}}/\protect\omega %
_{Bi}\simeq 0.13$, is shown by the black dot. The instability threshold
(dashed outer contour) encircles the range of unstable wavenumbers. The
parallel wavenumber $k_{z\protect\mu }V_{A}/\protect\omega _{Bi}$,
corresponding to the local maximum $\protect\gamma _{\protect\mu }/\protect%
\omega _{Bi}$, is shown by the blue dotted line. }
\end{figure}

The full wavenumber dependence of the CCOI dispersion (\ref{AW1}) is shown
in the contour plot Fig. 2 for the well above-threshold beam current $\bar{j}%
_{b}=0.3$ and thermal velocity $V_{Tb}/V_{A}=20$. It is seen that the
instability increment has a well-defined maximum in the $\left(
k_{z},k_{\perp }\right) $ plane. The instability boundary in the $\left(
k_{z},k_{\perp }\right) $ plane is shown in Fig. 2 by the outer dash line.
All wavenumbers inside the area below this line are unstable. It is seen
that the instability range is bounded in the $k_{\perp }$-space, and both
the lower and upper bounds, $\mu _{b1}$ and $\mu _{b2}$, are finite and
non--zero. In contrast, the lower bound of the unstable parallel wavenumber
range is zero. For the well over-threshold currents, the analytical
expressions for the boundaries in the $k$-space can be obtained as
\[
\mu _{b2}\simeq \bar{j}_{b}\frac{V_{Tb}}{V_{A}}\frac{k_{x}}{k}
\]%
for the high-$k_{\perp }$ boundary, and
\[
\mu _{b1}\simeq \frac{2}{3\mu _{b2}}
\]%
for the low-$k_{\perp }$ boundary.

Since the lower bound of unstable $k_{z}$ is zero, very long parallel wave
lengths can be generated by CCOI. However, there are several limitations at
small $k_{z}$ imposed by the finite field-aligned dimension of the system
and/or time scale of system variability. The latter limitation is related to
the fact that the instability growth time increases with decreasing $k_{z}$
and at some finite $k_{z}$ becomes longer than the characteristic evolution
time of the system.

\begin{figure}[tbp]
\label{f3} \plotone{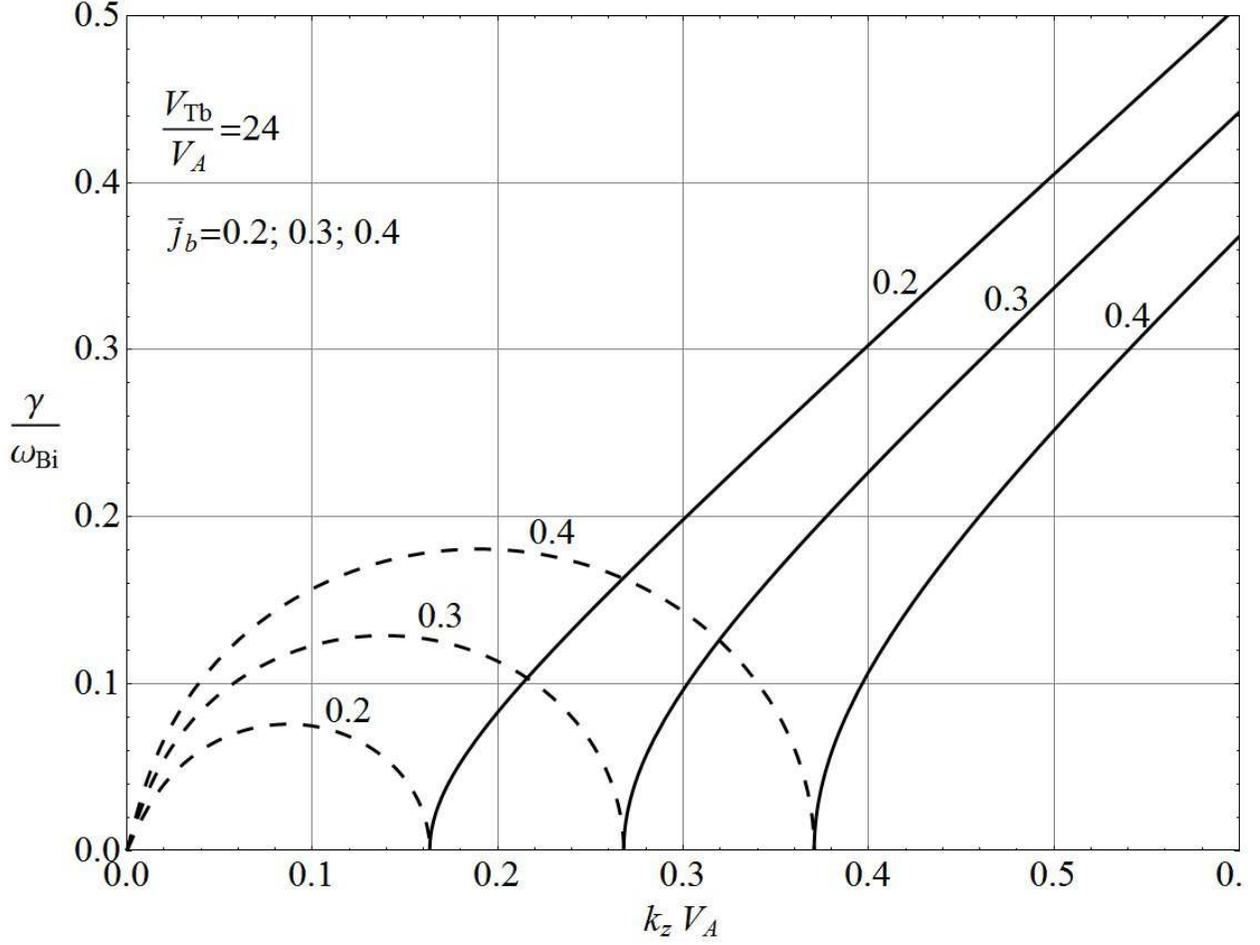}
\caption{CCOI growth rate (dash lines) as function of the normalized
parallel wavenumber $k_{z}V_{A}/\protect\omega _{Bi}$ for different beam
currents $\bar{j}_{b}=0.2$, $0.3$, and $0.4$ (for the lines from bottom to
top). The waves are aperiodically unstable at small $k_{z}V_{A}/\protect%
\omega _{Bi}$, and then become current-modified AWs at larger $k_{z}V_{A}/%
\protect\omega _{Bi}$ with real frequency (solid lines).}
\end{figure}

The $k_{z}$ dependence of the CCOI dispersion is shown for several values of
$\bar{j}_{b}$ in Fig. 3. For each $\bar{j}_{b}$ we fix $k_{\perp }$ at the
value where the CCOI increment $\gamma =\mathrm{Im}\left( \omega \right) $
passes through the absolute maximum. At small $k_{z}$ there is an unstable
wavenumber range where the mode is aperiodically growing. The increment
first increases linearly with $k_{z}$, then its increase slows down and
attains a maximum, and then decreases to zero. At larger $k_{z}$, above this
zero point, the CCOI dispersion becomes real and describes the usual Alfv%
\'{e}n wave with frequency shifted down by the compensated-current effects
(it is shown by the solid line in Fig. 3).

\begin{figure}[tbp]
\label{f4} \plotone{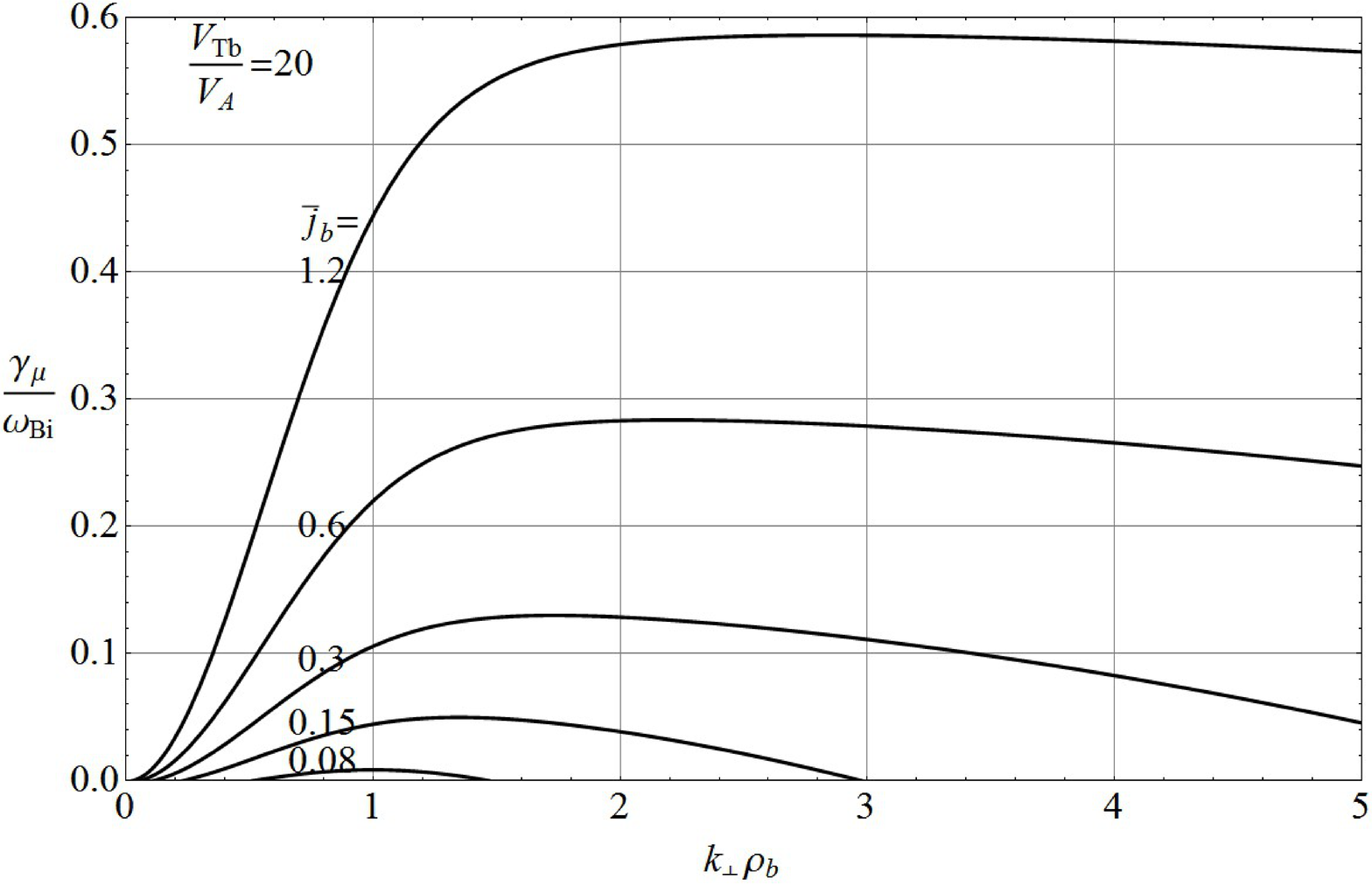}
\caption{CCOI growth rate at the local maximum $\protect\gamma _{\protect\mu %
}/\protect\omega _{Bi}$ as function of the normalized perpendicular
wavenumber $k_{\perp }\protect\rho _{b}$ for different beam currents $\bar{j}%
_{b}=1.2$, $0.6$ , $0.3$, $0.15$, and $0.08$ (for the lines from top to
bottom). The range of unstable wavenumbers extends with growing $\bar{j}_{b}$%
.}
\end{figure}

The easiest way to find the absolute maximum of (\ref{AW1}) is first to find
a local maximum of $\gamma $ with respect to $k_{z}$. This can be done
analytically, and for arbitrary $\mu _{b}$ we find
\begin{equation}
\frac{\gamma _{\mu }}{\omega _{Bi}}=\frac{1}{2}\left( \bar{j}_{b}\left(
1+A_{0}^{\prime }\left( \mu _{b}^{2}\right) \right) -\left( \frac{V_{A}}{%
V_{Tb}}\right) ^{2}\frac{\mu _{b}^{2}}{\bar{j}_{b}\left( 1+A_{0}^{\prime
}\left( \mu _{b}^{2}\right) \right) }\right) .  \label{gmu}
\end{equation}%
This maximum is attained at
\begin{equation}
\frac{k_{z\mu }V_{A}}{\omega _{Bi}}=\frac{1}{2}\sqrt{\bar{j}_{b}^{2}\left(
1+A_{0}^{\prime }\left( \mu _{b}^{2}\right) \right) ^{2}-\left( \frac{V_{A}}{%
V_{Tb}}\right) ^{4}\frac{\mu _{b}^{4}}{\bar{j}_{b}^{2}\left( 1+A_{0}^{\prime
}\left( \mu _{b}^{2}\right) \right) ^{2}}}.  \label{qmu}
\end{equation}

The perpendicular wavenumber dependence of the increment $\gamma _{\mu }$ is
shown in Fig. 4 for $V_{Tb}/V_{A}=20$ and for several values of the beam
current $\bar{j}_{b}=0.08$ (near-threshold value), $0.15$, $0.3$, $0.6$, and
$1.2$. It is seen that with growing $\bar{j}_{b}$ the perpendicular
wavenumber $\mu _{b\mathrm{m}}$, at which the instability increment attains
the absolute maximum $\gamma _{\mathrm{m}}$, increases. Also, the range of
unstable $\mu _{b}$ widens, such that smaller and larger perpendicular
wavenumbers are excited. This especially concerns higher $\mu _{b}>\mu _{b%
\mathrm{m}}$, where the increment is decreasing slowly and remains large,
comparable to the maximum value $\gamma _{\mathrm{m}}$. On the other hand,
in this high-$\mu _{b}$ range the finite gyroradius effects may become
important not only for the beam, but also for the background protons. The
finite-$\mu _{i}$ corrections that are neglected here will be investigated
in our forthcoming study.

The $\mu _{b}$-dependence of $k_{z\mu }$ (\ref{qmu}) is shown in Fig. 2 by
the blue dotted line. At certain point along this line, $\mu _{b}=\mu _{b%
\mathrm{m}}$ and $k_{z}=k_{z\mathrm{m}}$, the increment attains the absolute
maximum $\gamma _{\mu }=\gamma _{\mathrm{m}}$ for given plasma parameters.
The absolute maximum defines the instability growth rate. In general, the
normalized wavenumbers at maximum, $\mu _{b\mathrm{m}}$ and $k_{z\mathrm{m}%
}V_{A}/\omega _{Bi}$, depend on the beam current $\bar{j}_{b}$ and thermal
velocity $V_{Tb}/V_{A}$. In Fig. 2, where $\bar{j}_{b}=0.3$ and $%
V_{Tb}/V_{A}=20$, the maximum $\gamma _{\mathrm{m}}\simeq 0.13\omega _{Bi}$
is achieved with $\mu _{b\mathrm{m}}\simeq 1.73$ and $k_{z\mathrm{m}%
}V_{A}/\omega _{Bi}\simeq 0.14$. The same values can also be fond in Fig. 4
at the maximum of the curve for $\bar{j}_{b}=0.3$.

\subsection{Asymptotic scalings}

From (\ref{gmu}), it is possible to obtain two important scaling relations
for the instability growth rate $\gamma _{\mathrm{m}}$. In the
"near-threshold" regime, where $\bar{j}_{b}<3\bar{j}_{\mathrm{thr}}$, the
instability growth rate $\gamma _{\mathrm{m}}$ is proportional to the excess
of the beam current over the threshold one:
\begin{equation}
\gamma _{\mathrm{m}}\simeq 0.67\left( \bar{j}_{b}-\bar{j}_{\mathrm{thr}%
}\right) \omega _{Bi}.  \label{gmax1}
\end{equation}%
This maximum is achieved at $\mu _{b\mathrm{m}}\simeq 0.9$ and
\[
\frac{k_{z\mathrm{m}}V_{A}}{\omega _{Bi}}\simeq 0.34\sqrt{\bar{j}_{b}^{2}-%
\frac{\bar{j}_{\mathrm{thr}}^{4}}{\bar{j}_{b}^{2}}}
\]

Well above the threshold, $\bar{j}_{b}>3\bar{j}_{\mathrm{thr}}$, the maximum
growth rate, to the leading order, has the following linear scaling with the
beam current:
\begin{equation}
\gamma _{\mathrm{m}}\simeq 0.5\bar{j}_{b}\omega _{Bi}.  \label{gmax}
\end{equation}%
In this well over-threshold regime the dependence of $\mu _{b\mathrm{m}}$ on
the beam and plasma parameters can be approximated analytically by the
following expression:
\begin{equation}
\mu _{b\mathrm{m}}\simeq 0.8\left( \frac{V_{Tb}}{V_{A}}\bar{j}_{b}\right)
^{2/5}.  \label{mmax}
\end{equation}

The approximate small-$\bar{j}_{b}$ and large-$\bar{j}_{b}$ expressions (\ref%
{gmax1}) and (\ref{gmax}) connect smoothly in the intermediate range of
currents $\bar{j}_{b}\sim 3\bar{j}_{\mathrm{thr}}$, which means they can be
used for the growth rate estimations at all currents.

\subsection{Growth rate and inclination angle of CCOI}

The CCOI growth rate is determined by the absolute maximum of the
instability increment in the wavenumber space $\gamma _{\mathrm{m}}$. We
also define the instability wavenumber as the wavenumber $\mathbf{k}_{%
\mathrm{m}}=\left( k_{\perp \mathrm{m}},0,k_{z\mathrm{m}}\right) $ where the
instability increment attains the absolute maximum $\gamma _{\mathrm{m}}$.

\begin{figure}[tbp]
\label{f5} \plotone{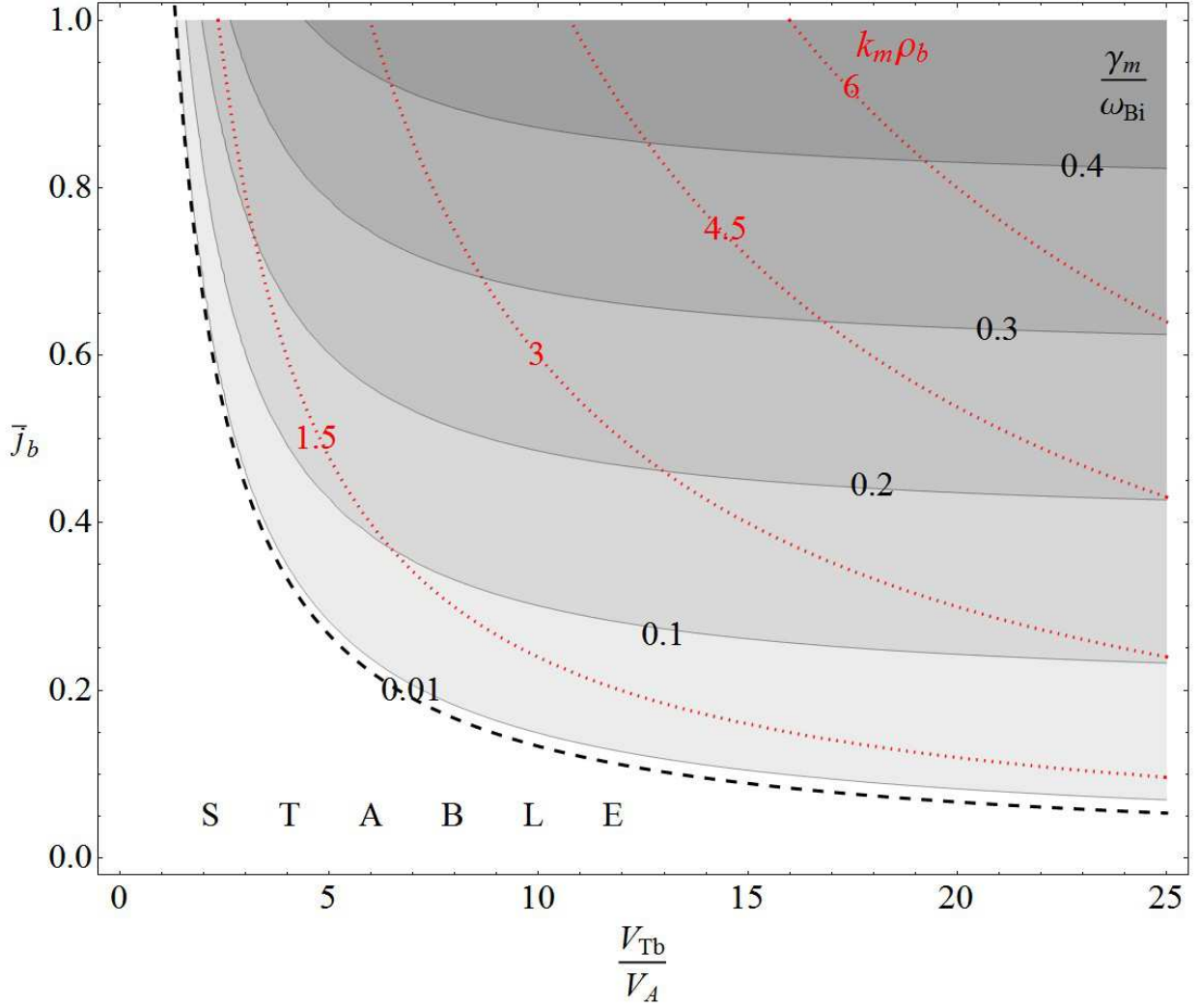}
\caption{(color online) Contour plot of the maximum CCOI growth rate $%
\protect\gamma _{\mathrm{m}}/\protect\omega _{Bi}$ (solid lines and shading)
as function of the beam current and thermal velocity. At large beam thermal
velocities $V_{Tb}/V_{A}>15$ CCOI growth rate is almost independent of $%
V_{Tb}/V_{A}$ (asymptotic instability regime). The contours of the
corresponding perpendicular wavenumber $k_{\mathrm{m}\perp }\protect\rho %
_{b} $ (dot red lines) are superimposed. Larger $k_{\mathrm{m}\perp }$ are
generated at larger beam current and temperature. The instability boundary
is shown by the dash line. }
\end{figure}

The contour plot of the CCOI growth rate $\gamma _{\mathrm{m}}$ is shown in
Fig. 5 as function of $\bar{j}_{b}$ (beam current in units of Alfv\'{e}n
current) and $V_{Tb}/V_{A}$ (beam thermal velocity in units of Alfv\'{e}n
velocity). Contours for $\gamma _{\mathrm{m}}$ are solid and emphasized by
the shadowing, such that the darker areas are more unstable. The full
normalized wavenumber $k_{\mathrm{m}}\rho _{b}$ of the most unstable
perturbations generated by CCOI is shown by dotted contours. In general, the
instability is stronger and generates larger wavenumbers at larger beam
currents and larger thermal velocities.

The instability threshold in the $\left( \bar{j}_{b},V_{Tb}\right) $ plane
is shown by the dash line, such that the range of beam currents and thermal
velocities above this line is CCOI-unstable. The threshold is very close to
the outer contour $\gamma _{\mathrm{m}}/\omega _{Bi}=0.01$. It is seen from
Fig. 5 that the increasing beam temperature favors the instability, making
the CCOI growth rate larger and the threshold current lower. In the wide
range of beam thermal velocities, $5<V_{Tb}/V_{A}<25$, the instability is
quite strong, $\gamma _{\mathrm{m}}\simeq \left( 0.1\div 0.2\right) \omega
_{Bi}$, with moderate beam currents $\bar{j}_{b}\sim 0.4$.

\begin{figure}[tbp]
\plotone{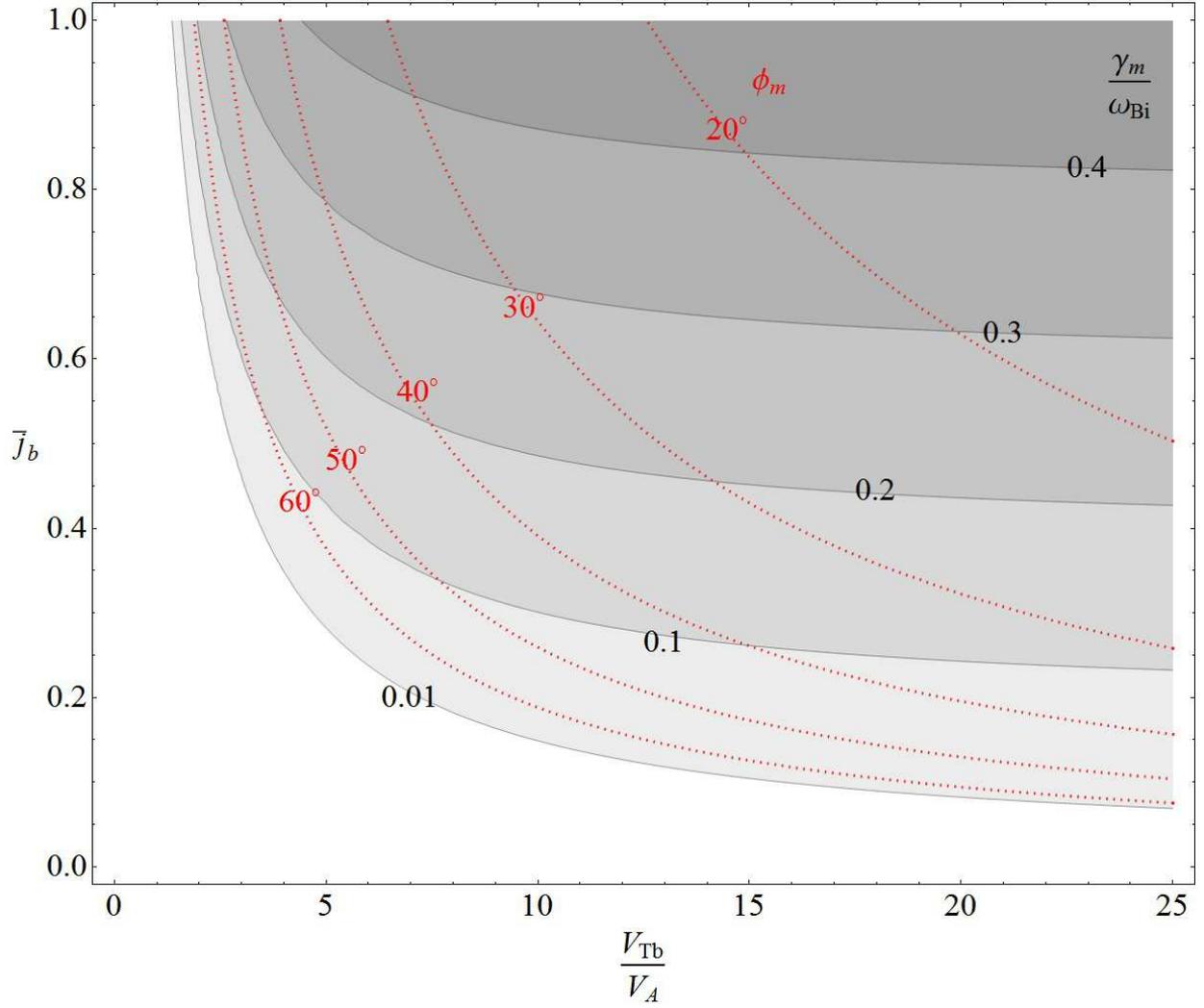}
\caption{(color online) The same as in Fig. 5, but with wavevector tilt angle $%
\protect\phi _{\mathrm{m}}$ (red dotted lines, values are in degrees)
superimposed on the CCOI growth rate $\protect\gamma _{\mathrm{m}}/\protect%
\omega _{Bi}$ (gray solid lines with shading). Less oblique waves are
generated by hotter beams.}
\end{figure}

It is interesting to estimate $\phi _{\mathrm{m}}$, the wavevector tilt angle of
most unstable perturbations with respect to the mean magnetic field.
This angle is given by
\begin{equation}
\mathrm{tg}\left( \phi _{\mathrm{m}}\right) =\frac{k_{\perp \mathrm{m}}}{k_{z%
\mathrm{m}}}=\frac{\mu _{b\mathrm{m}}}{k_{z\mathrm{m}}V_{A}/\omega _{Bi}}%
\frac{V_{A}}{V_{Tb}}.  \label{phimax}
\end{equation}%
$\allowbreak $In the described above case of $\bar{j}_{b}=0.3$ and $%
V_{Tb}/V_{A}=20$, it is $\phi _{\mathrm{m}}\simeq 32^{\circ }$. The
tilt angle $\phi _{\mathrm{m}}$ as function of $\bar{j}_{b}$ and $%
V_{Tb}/V_{A}$ is shown in Fig. 6. From this figure one can see that only
oblique fluctuations are generated by CCOI. In the near-threshold regime the
generated perturbations are very oblique, $45^{\circ }<\phi _{\mathrm{m}%
}<90^{\circ }$. However, with growing $\bar{j}_{b}$ and/or $V_{Tb}/V_{A}$,
the wavevector tilt angle quickly decreases and in the well over-threshold
regime the tilt angles are relatively small,
$\phi _{\mathrm{m}}\lesssim 20^{\circ }$.

In the near-threshold and well over-threshold regimes the tilt angle
can be estimated from the following scaling relations:
\begin{equation}
\left\{
\begin{array}{ll}
\mathrm{ctg}\left( \phi _{\mathrm{m}}\right) \simeq 0.38\sqrt{\bar{j}%
_{b}^{2}-\bar{j}_{\mathrm{thr}}^{2}\left( \frac{\bar{j}_{\mathrm{thr}}}{\bar{%
j}_{b}}\right) ^{2}} & \mathrm{for}\quad \bar{j}_{\mathrm{thr}}<\bar{j}_{b}<2%
\bar{j}_{\mathrm{thr}}; \\
\mathrm{ctg}\left( \phi _{\mathrm{m}}\right) \simeq 0.76\left( \frac{\bar{j}%
_{b}}{\bar{j}_{\mathrm{thr}}}\right) ^{3/5} & \mathrm{for}\quad 3\bar{j}_{%
\mathrm{thr}}\leq \bar{j}_{b}<2.%
\end{array}%
\right.  \label{phimax1}
\end{equation}

\section{CCOI in the solar wind upstream of the terrestrial bow
shock}

As a possible example where CCOI can develop we consider compensated-current
systems created by proton beams propagating upstream of the terrestrial bow
shock. The observed proton beams can be categorized into three classes
(Paschmann et al., 1981; Tsurutani \& Rodriguez 1981): (1) fast beams with
temperatures 10$^{6}$ $<T_{b}<$ 10$^{7}$ $^{\circ }$K and velocities (in the
solar wind frame) $10$ $<V_{b}/V_{A}<$ $25$ created by the protons
"reflected" from the quasi-perpendicular shocks; (2) slow hot beams with 5$%
\cdot $10$^{7}$ $<T_{b}\lesssim $ 10$^{8}$ $^{\circ }$K and 1 $<V_{b}/V_{A}<$
7 created by "diffuse" protons upstream of the quasi-parallel shocks; and
(3) "intermediate" beams with $T_{b}\gtrsim $ 10$^{7}$ $^{\circ }$K and $%
V_{b}/V_{A}\sim $ $10$ observed in the regions in-between. The beam number
densities are similar for all three classes and vary in the range $%
n_{b}/n_{0}=$ $0.01\div 0.1$.

1. The "reflected" beams propagate from quasi-perpendicular shocks with high
velocity $V_{b}/V_{A}=$ 15 (which with $n_{b}/n_{0}=0.03$ gives $\bar{j}%
_{b}=0.45$) and $V_{Tb}/V_{A}=5$. With these "representative" values we
estimate the growth rate $\gamma _{\mathrm{m}}/\omega _{Bi}\simeq 0.08$
attained at $\mu _{b\mathrm{m}}=1.1$, $k_{z\mathrm{m}}V_{A}/\omega
_{Bi}=0.14 $. The tilt angle $\phi _{\mathrm{m}}\simeq $ $57^{\circ }$%
.

2. For the "intermediate" beams with $V_{b}/V_{A}=$ 10, $n_{b}/n_{0}=0.03$, $%
\bar{j}_{b}=0.3$, and $V_{Tb}/V_{A}=10$ the CCOI growth rate is quite large,
$\gamma _{\mathrm{m}}/\omega _{Bi}\simeq 0.1$, and is attained at $\mu _{b%
\mathrm{m}}=1.35$, $k_{z\mathrm{m}}V_{A}/\omega _{Bi}=0.13$. The
corresponding tilt angle $\phi _{\mathrm{m}}\simeq $ $46^{\circ }$.

3. The "diffuse" beams propagate along magnetic fields linked to the regions
where the shock is quasi-parallel (the angle between the shock normal and
magnetic field $\mathbf{B}_{0}$ is less than 45$^{\circ }$). Using the
"representative" values for these beams, $n_{b}/n_{0}=0.03$, $V_{b}/V_{A}=$
5 (hence $\bar{j}_{b}=0.15$), and $V_{Tb}/V_{A}=15$, we find the CCOI growth
rate $\gamma _{\mathrm{m}}/\omega _{Bi}\simeq 0.04$. The characteristic
wavenumbers of excited waves are $\mu _{b\mathrm{m}}=1.2$, $k_{z\mathrm{m}%
}V_{A}/\omega _{Bi}=0.06$, such that the angle between $\mathbf{k}_{\mathrm{m%
}}$ and $\mathbf{B}_{0}$ is about $\phi _{\mathrm{m}}\simeq $ $54^{\circ }$.

As is seen from the above estimations, the "intermediate" beams are most
favorable for CCOI. With larger beam currents and/or thermal velocities the
CCOI growth rate is larger and the tilt angle is smaller (see Fig. 6
for specific numbers). Say, for the "intermediate" beam with $V_{b}/V_{A}=$ $%
10$, $V_{Tb}/V_{A}=10$ and elevated density $n_{b}/n_{0}=0.09$ the
instability is very strong, $\gamma _{\mathrm{m}}/\omega _{Bi}\simeq 0.4$,
and the tilt angle $\phi _{\mathrm{m}}\simeq $ $20^{\circ }$.

Oblique waves with $\phi _{\mathrm{m}}$ scattered between
$0^{\circ }$ and $180^{\circ }$ were observed by Cluster
in the foreshock, with the dominant wave fraction concentrated in the range
$\phi _{\mathrm{m}}\simeq 10-30^{\circ }$ and several minor peaks
(see Fig. 13 by Narita et al., 2006).\ Two peaks at different $\phi _{%
\mathrm{m}}$ were also observed by Hobara et al. (2007) for 30-s
waves. In addition to the main peak, Narita et al. (2006) also found
minor fractions of more oblique waves in the quasi-parallel foreshock:
forward fraction at $\phi _{\mathrm{m}}\simeq 40-70^{\circ }$, and
backward fractions at $\phi _{\mathrm{m}}\thicksim 140^{\circ }$
and $\phi _{\mathrm{m}}\thicksim 170^{\circ }$ . In the
quasi-perpendicular foreshock the sub-dominant wave fractions at $\phi _{%
\mathrm{m}}\thicksim 60^{\circ }$ and $\phi _{\mathrm{m}}\lesssim
90^{\circ }$ are relatively larger than in the quasi-parallel
foreshock. The wave mode composition of these spectra, especially in the
quasi-perpendicular foreshock (see Narita et al., 2006), is uncertain.
Multiple spectral peaks and variable properties of observed waves suggest
that several modes and instabilities contribute to the foreshock wave
spectra.

To generate the dominant wave fraction at $\phi _{\mathrm{m}}\simeq
10-30^{\circ }$, CCOI needs large enough $\bar{j}_{b}$
and/or $V_{Tb}/V_{A}$ from the range above the red line
corresponding to $30^{\circ }$ in Fig. 6. Less restrictive
conditions are required for the generation of sub-dominant wave spectra in
quasi-perpendicular foreshocks (values of $\bar{j}_{b}$ and $%
V_{Tb}/V_{A}$ along the $60^{\circ }$ line in Fig. 6 for
the peak at $\phi _{\mathrm{m}}\thicksim 60^{\circ }$, and just
above the dash line for the very oblique waves at $\phi _{\mathrm{m}%
}\lesssim 90^{\circ }$).

If we turn now to the full wavenumber distributions of observed
waves (Fig. 9 by Narita et al.), we find a quite different picture with two
about equal peaks that do not map onto unequal peaks on the angle
distributions shown in Fig. 13 by Narita et al. The peak at $k_{\mathrm{m}%
}\rho _{i}\sim 0.4$ in the quasi-perpendicular foreshock can be
easily generated by CCOI with relatively small values of $\bar{j}_{b}$
and $V_{Tb}/V_{A}$ around the line $k_{\mathrm{m}}\rho
_{b}=$ $k_{\mathrm{m}}\rho _{i}\sqrt{T_{b}/T_{i}}\simeq $
$1.5$ in Fig. 5). Another peak\ in the quasi-perpendicular
foreshock at smaller $k_{\mathrm{m}}\rho _{i}\approx 0.06$ is more
difficult to reproduce by CCOI, which would require quite high values of $%
T_{b}/T_{i}>10^{2}$ more typical for the quasi-parallel foreshock.
In the quasi-parallel foreshock, both peaks can be generated by CCOI under
reasonable beam and plasma conditions.

There is an apparent contradiction between CCOI properties and
properties of foreshock waves observed by Narita et al. (2006). Namely,
Narita et al. estimated the real wave frequency in the plasma frame $\sim
0.2\omega _{Bi}$, which contradicts the aperiodic nature of
perturbations generated by CCOI. One should however note that in a
non-stationary plasma with time-varying parameters the wave frequency
and polarization are not conserved quantities but evolve in time (see e.g.
Mendon\c{c}a, 2009; Lade et al., 2011, and references therein). Therefore, in
the highly dynamic foreshock environment the aperiodic waves generated by CCOI can
develop significant real frequencies and contribute to the wave spectra
observed by Narita et al. (2006). For example, from Fig. 3 it follows that
the aperiodic waves generated by CCOI when the current was $\bar{j}_{b}=0.4$%
possess the parallel wavenumber $k_{z}V_{A}\approx 0.2\omega _{Bi}$%
, which, being conserved in time, rises the wave frequency to $%
\sim 0.1\omega _{Bi}$ when the time-varying
$\bar{j}_{b}$ drops below $0.2$ and to $\sim 0.2\omega _{Bi}$
when $\bar{j}_{b}$ drops below $0.1$.
The wave phase velocity also varies in the time-varying conditions,
which makes the wave mode identification based on the phase velocity histograms
(Figs. 10-12 by Narita et al., 2006) even more uncertain.
Particular regimes of the wave temporal evolution are studied
by far insufficiently and further investigations are needed for this
process in the foreshock conditions.

Since the stable range is bounded by the threshold current, $\bar{j}_{b}=$ $%
\bar{j}_{thr}$, and the threshold current $\bar{j}_{thr}$ depends on $%
V_{Tb}/V_{A}$ (\ref{jthr}), the measured values of $\bar{j}_{b}$ and $%
V_{Tb}/V_{A}$ should be statistically constrained. Namely, if the boundary
in the scatter plot of measured values ($\bar{j}_{b}$, $V_{Tb}/V_{A}$) can
be approximated analytically as
\begin{equation}
\bar{j}_{b}\frac{V_{Tb}}{V_{A}}\simeq a_{CCOI},  \label{CCI}
\end{equation}%
with $a_{CCOI}=1.5\div 2$, that would suggest that the beam currents and/or
temperature are regulated by CCOI. We are not aware of such correlation
measurements in the terrestrial foreshock.

The observed satellite-frame frequency is determined by the large Doppler
shift and can be estimated as $\omega _{\mathrm{sat}}\simeq $ $k_{\perp }V_{%
\mathrm{SW}}\mathrm{sin}\theta _{\mathrm{VB}}\simeq $ $\omega
_{Bi}(V_{sw}/V_{Tb})\mu _{b\mathrm{m}}\mathrm{sin}\theta _{\mathrm{VB}}$,
where $\theta _{\mathrm{VB}}$ is the angle between the solar wind speed and $%
\mathbf{B}_{0}$. Having in mind that $\mu _{b\mathrm{MAX}}=0.9\div 2$, with
the same "representative" values as above, and $\theta _{\mathrm{VB}}\simeq
\pi /4$, we obtain $\omega _{\mathrm{sat}}\simeq \left( 0.1\div 1\right)
\omega _{Bi}$. The waves in this frequency range are regularly observed by
satellites.

In the time intervals when $\theta _{\mathrm{VB}}\simeq 0$, the
satellite-frame frequencies of the CCOI fluctuations are determined by the
(smaller) parallel wavenumbers and $\omega _{\mathrm{sat}}$ reduces to $\sim
0.05\omega _{Bi}$. Consequently, we have another expected observational
signature of CCOI: the measured wave energy in the low-frequency band $%
\left( 0.01\div 0.1\right) \omega _{Bi}$ should be larger in the cases $%
\theta _{\mathrm{VB}}\simeq 0$ than in the cases $\theta _{\mathrm{VB}%
}\gtrsim \pi /4$. We are not aware if such a trend is observed.

\section{Discussion}

To understand the physical nature of CCOI we note that the destabilizing
term proportional to $\bar{j}_{b}^{2}$\ in the dispersion equation (\ref{DE}%
) comes from the product of non-diagonal elements $\varepsilon _{xy}$\ and $%
\varepsilon _{yx}=-\varepsilon _{xy}$\ of the dielectric tensor (\ref{e}),
which are dominated by the background electron current $j_{e}$\ and the
reduced beam current $A_{0}^{\prime }\left( \mu _{b}^{2}\right) j_{b}$:\
\begin{equation}
\varepsilon _{xy}=\varepsilon _{xy}^{\left( e\right) }+\varepsilon
_{yx}^{\left( b\right) }\simeq i\left( \frac{4\pi }{B_{0}}\right) \left(
\frac{k_{z}c}{\omega ^{2}}\right) \left[ n_{0}eV_{e}+n_{b}eV_{b}A_{0}^{%
\prime }\left( \mu _{b}^{2}\right) \right] \simeq i\frac{\omega _{Pp}}{%
\omega }\frac{k_{z}c}{\omega }\left[ 1+A_{0}^{\prime }\left( \mu
_{b}^{2}\right) \right] \bar{j}_{b}.
\end{equation}%
\ The zero net current condition $n_{0}eV_{e}=n_{b}eV_{b}$\ is used in the
above expression.

From the electron contribution $\varepsilon _{xy}^{\left( e\right) }$,\ and
the beam contribution $\varepsilon _{yx}^{\left( b\right) }$, we see that
the fluctuating electron and beam ion currents can be expressed via
fluctuating magnetic field $\delta \mathbf{B}_{\perp }$\ as $\delta \mathbf{j%
}_{e\perp }\simeq j_{e}\delta \mathbf{B}_{\perp }/B_{0}$ and $\delta \mathbf{%
j}_{b\perp }\simeq A_{0}^{\prime }\left( \mu _{b}^{2}\right) j_{b}\delta
\mathbf{B}_{\perp }/B_{0}$, respectively. \ These first-order currents have
the following simple interpretation. The frozen-in electron current, flowing
along the curved field lines, $\mathbf{B}=\mathbf{B}_{0}+\delta \mathbf{B}%
_{\perp }$, deviates in the $\delta \mathbf{B}_{\perp }$\ direction thus
developing a perpendicular component $\delta j_{e\perp }/j_{e}=\delta
B_{\perp }/B_{0}$. On the contrary, the ion beam current is partially
unfrozen by the large gyroradius of the beam ions, which reduces $\delta
j_{b\perp }$\ by the factor $A_{0}^{\prime }\left( \mu _{b}^{2}\right) $. As
a result, even if the zero-order currents $j_{e}$\ and $j_{b}$\ are
compensated ($j_{b}+j_{e}=$\ $0$), they induce the first-order electron\ and
beam ion currents\ that are not compensated, $\delta \mathbf{j}_{e\perp }+$ $%
\delta \mathbf{j}_{b\perp }\neq 0$. The resulting first-order net current $%
\delta \mathbf{j}_{\perp }\simeq \left( 1+A_{0}^{\prime }\left( \mu
_{b}^{2}\right) \right) j_{b}\delta \mathbf{B}_{\perp }/B_{0}$ makes AWs
aperiodically unstable. CCOI is therefore the current-driven instability in
two respects: (1) the instability source is the compensated zero-order
currents, which generate (2) the uncompensated first-order current $\delta
j_{\perp }$\ responsible for the instability.

As follows from the above explanation, the physical nature of CCOI is
different from that of the fire-hose instabilities, including parallel and
oblique fire-hoses driven by the effective anisotropic plasma pressure $%
T_{\parallel }>T_{\perp }$ (see e.g. Hellinger \& Matsumoto, 2000).

Given its non-resonant driving mechanism, CCOI depends on the bulk
parameters of plasma species rather than on the local behavior of their
velocity distributions. By using other velocity distributions instead of
Maxwellian, one would obtain similar results with the destabilizing factor
proportional to $j_{b}$, but with another demagnetization function replacing
$A_{0}^{\prime }\left( \mu _{b}^{2}\right) $. The behavior of any particular
demagnetization function is expected to be as regular as $A_{0}^{\prime
}\left( \mu _{b}^{2}\right) $, with the same limits $\rightarrow -1$\ at $%
k_{\perp }\rightarrow 0$\ and $\rightarrow 0$\ at $k_{\perp }\rightarrow
\infty $. The instability is therefore expected to be quite robust and,
contrary to the resonant current-driven instabilities, not suffering from
the fast saturation by the local plateau formation in the particle velocity
distributions.

It is interesting to note that the growth rate of the Winske-Leroy
instability in the well over-threshold regime (their formula (16)) can be
expressed in terms of the beam current as $\gamma _{\mathrm{WL}}\simeq 0.5%
\bar{j}_{b}\omega _{Bi}$, which is exactly the same scaling as for CCOI (\ref%
{gmax}). The same scaling suggests that both instabilities are driven by the
same factor. Winske \& Leroy have stressed that their strong non-resonant
instability is not of the fire-hose type, but did not explain its physical
nature. After inspecting derivations by Winske \& Leroy (1983), we found
that their most unstable regime (equations (14)-(16) in their paper) is
indeed driven by the same factor as CCOI: the non-compensated wave currents
developed in response to the compensated global currents.

A similar compensated-current instability of MHD-like modes has been found
recently by Bell (2004, 2005). The Bell instability arises in response to
the currents induced by cosmic rays around super-nova remnants. Again, using
equation (5) by Bell (2005), it is easy to see that this instability has the
maximum $\gamma _{\mathrm{Bell}}\simeq 0.5\bar{j}_{b}\omega _{Bi}$, attained
at $\left\vert k_{z\mathrm{m}}\right\vert V_{A}/\omega _{Bi}=$ $0.5\bar{j}%
_{b}$. These expressions are exactly the same as for the Winske-Leroy
instability. Also, similarly to the Winske-Leroy instability, the Bell
instability maximizes at parallel propagation, $k_{\perp }=0$. Bell's
analysis differs from Winske-Leroy's one in how the background plasma, beam,
and unstable modes are treated. Winske \& Leroy (1983) used a fully kinetic
theory assuming a shifted Maxwellian proton beam, whereas Bell (2004)
reduced the problem to the "hybrid" MHD-kinetic one (MHD with the currents
calculated kinetically), and used a power-law momentum distribution of the
beam ions.

Both the Winske-Leroy and the Bell instabilities arise because of
essentially the same physical effect: suppression of the wave response to
the beam ion current by the large factor $k_{z}V_{bz}/\omega _{Bi}>1$
(parallel dispersion effect). The wave response to the background electron
current survives such suppression because the electrons with small $%
k_{z}V_{ez}/\omega _{Be}<1$ remain magnetized. This implies a physical
interpretation somehow different from that proposed by Bell (2004, 2005),
who described it in terms of large ion gyroradius (\emph{perpendicular} ion
scale). In our opinion, the reducing factor in this case is the large \emph{%
parallel} scale of the beam ions, $\lambda _{bz}=V_{bz}/\omega _{Bi}$, which
defines the ion-cyclotron time-of-flight distance along the background
magnetic field. If the \emph{parallel} wavelength is comparable or shorter
than this distance, $k_{z}\lambda _{bz}>1$, the wave response to the beam
current is reduced by the effective ion demagnetization. Both the
Winske-Leroy and Bell instabilities are strongest at parallel propagation
and are physically the same instability that can be named a
compensated-current parallel instability (CCPI).

Effects of the large ion gyroradius in the beam, which are determined by the
\emph{perpendicular} ion motion and \emph{perpendicular} wavenumber
dispersion (factor $k_{\perp }\rho _{b\perp }=$ $k_{\perp }V_{Tb\perp
}/\omega _{Bi}$), are considered in the present paper. Similarly to
CCPI, the wave response to the beam ion current is also suppressed in CCOI,
but the nature of this suppression is different. Contrary to CCPI, for which
the parallel dispersive effects of finite $k_{z}\lambda _{bz}$ are
important, CCOI is caused by the perpendicular dispersive effect of finite
$k_{\perp }\rho _{b\perp }$. Consequently, CCOI develops in quite
different wavenumber range characterized by large $k_{\perp }$.

Concluding above comparisons, we expect several peaks in the spectrum of
unstable fluctuations in the compensated-current systems with ion beams. In
particular, two current-driven instabilities arise in the case of hot and
fast ion beams: one at parallel propagation $\phi =0$ (i.e. at $k_{\perp }=0$%
) for CCPI studied by Winske \& Leroy (1983) and Bell (2004), and another at
$\phi =\phi _{\mathrm{m}}$ defined by (\ref{phimax}) (i.e. at $k_{\perp
}=k_{\perp \mathrm{m}}$) for CCOI studied here. Because of the same
destabilizing factor $\bar{j}_{b}$, CCPI and CCOI have the same asymptotic
scaling $\gamma _{\mathrm{CCPI}}\sim \gamma _{\mathrm{CCOI}}\sim $ $0.5\bar{j%
}_{b}\omega _{Bi}$ at large $\bar{j}_{b}\gg \bar{j}_{\mathrm{thr}}$.
However, in the cases where $\bar{j}_{b}$ is not much larger than $\bar{j}_{%
\mathrm{thr}}$, the additional degree of freedom $k_{\perp }\rho _{b\perp
}\neq 0$ makes CCOI more flexible in finding larger growth rates as compared
to CCPI.

For hot and fast ion beams ($V_{Tb}\geq V_{b}\gg V_{A}$) the ion two-stream
and Buneman instabilities are inefficient, and the main competitors of CCOI
are the mentioned above left- and right-hand polarized resonant
instabilities (Gary, 1985) and the non-resonant instabilities studied by
Sentmann et al. (1981), Winske \& Leroy (1983), and Bell (2004). For
example, from the upper curve in Fig. 3b by Gary (1985) for the left-hand
polarized instability driven by the beam $n_{b}/n_{0}=0.02$ and $%
V_{b}/V_{A}=10$, we find the growth rate $\gamma _{\mathrm{LH}}/\omega
_{Bi}\simeq 0.07$. For the same plasma parameters, the CCOI growth rate is
somehow larger, $\gamma _{\mathrm{m}}/\omega _{Bi}\simeq 0.08$. This value
is also larger than that for the Winske-Leroy nonresonant instability at the
same parameters. Since the differences are not large, all these
instabilities can compete in the typical foreshock conditions.

At lower beam velocities, $V_{A}\leq V_{b}\leq 2V_{A}$, and lower beam
temperatures, $T_{b}\sim T_{p}$,
resonant instabilities of the parallel fast and oblique Alfv\'{e}n modes
are stronger than the parallel ones (Voitenko, 1998; Daughton et al.
1999; Gary et al. 2000; Voitenko \& Goossens, 2003; Verscharen \&
Chandran, 2013) and can compete with CCOI. Since the CCOI theory for this
parameter range is not yet developed, the quantitative comparison of CCOI
with these instabilities is postponed for a future study.

\section{Conclusions}

We found a new oblique Alfv\'{e}nic instability, CCOI, driven by compensated
currents flowing along the mean magnetic field. The instability arises on
the Alfv\'{e}n mode dispersion branch due to the coupling to the fast mode
via the current term proportional to $\bar{j}_{b}$. The instability is
enforced by the increasing current $\bar{j}_{b}$ and beam thermal spread $%
V_{Tb}$.

The physical mechanism of this instability is as follows. Because of the
finite ion gyroradius effects, the oblique Alfv\'{e}nic perturbations react
differently on the current carried by the beam ions and the current carried
by the electrons. Namely, the wave response to the beam ion current is
reduced by the averaging over the large ion gyroradius, whereas the small
electron gyroradius leave the electron current response practically
unaffected. Ultimately, the difference between the electron and ion
responses results in the net first-order current that shifts the Alfv\'{e}n
wave frequency squared below zero making the wave aperiodically unstable.

Our results show that in many astrophysical and space plasma settings,
comprising ion beams and return electron currents, the CCOI is a strong
competitor for the CCPIs studied by Winske \& Leroy (1984) and Bell (2004;2005),
as well as for the beam-driven firehose and kinetic instabilities.

The main CCOI properties are:

1. The instability is driven by the perpendicular dispersive effects of
finite $k_{\perp }\rho _{b\perp }$, which result in the uncompensated wave
currents developed in response to the compensated zero-order currents.

2. The threshold beam current is $\bar{j}_{\mathrm{thr}}\simeq
1.33V_{A}/V_{Tb}$, and the instability growth rate (maximal increment) is
\begin{equation}
\left\{
\begin{array}{ll}
\gamma _{\mathrm{m}}\simeq 0.67\left( \bar{j}_{b}-\bar{j}_{\mathrm{thr}%
}\right) \omega _{Bi}, & \quad \mathrm{for}\quad \bar{j}_{\mathrm{thr}}<\bar{%
j}_{b}<2\bar{j}_{\mathrm{thr}}; \\
\gamma _{\mathrm{m}}\simeq 0.5\bar{j}_{b}\omega _{Bi}, & \quad \mathrm{for}%
\quad 2\bar{j}_{\mathrm{thr}}\leq \bar{j}_{b}<2.%
\end{array}%
\right.  \label{max}
\end{equation}%
The upper bound on $\bar{j}_{b}$ appears here because of our initial
approximations, which restrict the range of tractable beam currents to $\bar{%
j}_{b}<2$ (this follows from the low-frequency approximation used, $\gamma _{%
\mathrm{m}}^{2}<\omega _{Bi}^{2}$).

3. The approximate expressions (\ref{max}) are valid in the range $%
1.33V_{A}/V_{Tb}<\bar{j}_{b}<2$, which is not empty if the beam temperature
is sufficiently high to make $V_{Tb}/V_{A}>0.7$. The $k_{\perp }$- dependent
growth rate $\gamma _{\mu }$ (\ref{gmu}) is valid for stronger currents $%
\bar{j}_{b}>2$, but only in the wavenumber ranges where $\gamma _{\mu
}<\omega _{Bi}$.

4. The range of unstable perpendicular wavenumbers is narrow in the
near-threshold regime, but expands as the beam current grows. Consequently,
a wide-band wave spectrum can be generated well above the threshold.

5. We found that the optimal perpendicular wavenumber for the instability is
$k_{\perp }\rho _{b}\gtrsim 1$ and the instability is very strong, $\gamma _{%
\mathrm{m}}\simeq \left( 0.1\div 0.5\right) \omega _{Bi}$ for reasonable
beam currents $\bar{j}_{b}\simeq 0.1\div 1$.

6. An essential characteristics of the fluctuations generated by CCOI is
their obliquity. In the near-threshold regime $\bar{j}_{\mathrm{thr}}<\bar{j}%
_{b}<2\bar{j}_{\mathrm{thr}}$ the generated fluctuations are very oblique, $%
50^{\circ }<\phi _{\mathrm{m}}<90^{\circ }$. Well above the threshold, the
instability becomes less oblique and $\phi _{\mathrm{m}}$ can drop below $%
20^{\circ }$ for strong enough beam currents.

Such oblique fluctuations, regularly registered in the terrestrial
foreshock, can be explained by CCOI. Other competing instabilities, like
left-/right-hand resonant (Gary, 1985), fire-hose (Sentmann et al., 1981),
"anti-parallel non-resonant" (Winske \& Leroy, 1983), and "parallel
non-resonant" (Bell, 2004,2005) are magnetic field-aligned and hence cannot
explain oblique fluctuations.

\acknowledgments

This research was supported by the Belgian Science Policy Office (through
Prodex/Cluster PEA 90316 and IAP Programme project P7/08 CHARM), and by the
European Commission (through FP7 Program project 313038 STORM).

\end{document}